\newcommand\pa{\partial}

\newcommand{\beque}{\begin{equation*}}
\newcommand{\eeq}{\end{equation}}
\newcommand{\beq}{\begin{equation}}
\newcommand{\eeque}{\end{equation*}}
\newcommand{\beqnl}{\begin{eqnarray}}
\newcommand{\eeqna}{\end{eqnarray*}}
\newcommand{\beqna}{\begin{eqnarray*}}
\newcommand{\eeqnl}{\end{eqnarray}}

\newcommand\lagr{{\mathcal L}}
\newcommand\ham{{\mathcal H}}

\documentclass[10pt]{iopart}
\usepackage{amssymb}
\usepackage{amsfonts}
\usepackage{amsbsy}
\usepackage{graphicx}
\usepackage{color}

\begin{document}


\title{The Hopfield model revisited: Covariance and Quantization}

\author{F.~Belgiorno$^{1,2}$, S.L.~Cacciatori$^{3,4}$, F.~Dalla~Piazza$^{5}$}

\address{$^1$ Dipartimento di Matematica, Politecnico di Milano, Piazza Leonardo 32, IT-20133
Milano, Italy}
\address{$^2$ and INdAM-GNFM}
\address{$^3$ Department of Science and High Technology, Universit\`a dell'Insubria, Via Valleggio 11, IT-22100 Como, Italy}
\address{$^4$ INFN sezione di Milano, via Celoria 16, IT-20133 Milano, Italy}
\address{$^5$ Universit\`a ``La Sapienza'', Dipartimento di Matematica, Piazzale A. Moro 2, I-00185, Roma,   Italy}

\begin{abstract}
There are several possible applications of quantum electrodynamics in dielectric media which 
require a  quantum description for the electromagnetic field interacting with matter fields. 
The associated quantum models can refer to macroscopic electromagnetic 
fields or, in alternative, to mesoscopic fields (polarization fields) describing an effective interaction between 
electromagnetic field and matter fields. 
We adopt the latter approach, and 
focus on the Hopfield model for the electromagnetic field in a dielectric
dispersive medium in a framework in which space-time dependent mesoscopic 
parameters occur, like susceptibility, matter resonance frequency, and also coupling 
between electromagnetic field and polarization field. Our most direct goal is to 
describe in a phenomenological way a space-time varying dielectric perturbation
induced by means of the Kerr effect in nonlinear dielectric media.
This extension of the model is implemented by means of a Lorentz-invariant
Lagrangian which, for constant microscopic parameters, and in the rest frame, 
coincides with the standard one. Moreover, we deduce a
covariant scalar product and  provide a canonical quantization scheme which keeps into account
the constraints implicit in the model. Examples of viable applications are indicated.
\end{abstract}

\submitto{Physica Scripta}

\vskip2pc
\noindent

\maketitle

\ioptwocol


\section{Introduction}

A longstanding field of investigation for quantum field theory is represented by pair creation in external fields or 
by moving boundaries. We limit ourselves to quote a couple of seminal {\color{black} papers which were published} at the birth of 
modern quantum field theory \cite{heisenberg,schwinger}. 
We are mainly interested in photon pair creation associated with variations of the dielectric 
constant in a dielectric medium, which has been a subject of important investigations, 
as e.g. a series of papers by Schwinger concerning a possible relation between 
dynamical Casimir effect and sonoluminescence \cite{schwinger-sono}. 
In this paper, instead of quantizing phenomenologically the electromagnetic field in presence 
of a dielectric medium (see e.g. \cite{luks}),  we delve into a less phenomenological situation in which dielectric properties 
are modeled as in the Hopfield model:  the electromagnetic field interacts with a set of oscillators reproducing sources for dispersive properties 
of the electromagnetic field in matter \cite{hopfield,fano,kittel,davydov}. We refer to a more general situation 
where electric susceptibility, resonance frequencies of the electromagnetic field and also the coupling between 
electromagnetic field and oscillators (the latter ones can be identified with the mesoscopic polarization fields) depend on 
space-time variables. In this respect, we can refer our approach to other models which generalize the Hopfield model, 
see \cite{barnett,suttorp-epl,suttorp-jpa}, but {\color{black} we stress that we don't take into account absorption in our paper}. The latter assumption is reasonable as far as emission phenomena we are 
interested in are not too near the absorption region, and the focus is pair-creation. Any framework including absorption would imply 
a much more tricky approach (cf. e.g. \cite{barnett,suttorp-epl,suttorp-jpa}).\\

In order to corroborate the physical interest of our model and its canonical quantization, 
we recall that, by means of the Kerr effect,  it is possible to induce 
dielectric perturbations which propagate in the dielectric medium. These 
perturbations represent inhomogeneities which arise as a consequence of intense laser pulse propagating in a dielectric
medium (see e.g. \cite{boyd}). As a
result of microscopic interactions, involved in non-linear electrodynamics, 
 a mean motion of a dielectric perturbation with a different refractive index occurs. 
Instead of attempting immediately a first principle study on this subject, i.e. instead of
working out a microscopic model for this case, we begin considering a semi-phenomenological
approach, in which we adopt a Lagrangian model which would be `first principle-based' apart for
the appearance in the Lagrangian of a (phenomenological) contribution to the refractive index
arising from the Kerr effect. Interesting results can be still 
deduced from this framework, as well as a well-defined route for quantizing the model.
Our aim is to get a model which allows to deal with sufficiently general situations
of interest, as e.g. the ones created by means of the Kerr effect: traveling dielectric perturbations 
moving with different laws of motions should be allowed. For example, a uniformly moving perturbation 
(which is characteristic of the Kerr effect), but also an accelerating one
and even a rotating one, each of which represents a very interesting benchmark
for photon pair creation in external fields (or under changing external conditions). 
For situations covering the uniformly moving case, its relation with 
analogue gravity framework and for experimental measure of Hawking radiation, see e.g. 
\cite{philbin-leonhardt,belgiorno-prl,rubino-njp,belgiorno-prd,petev-prl,finazzi-carusotto-pra,finazzi-carusotto-pra14}.
Thus our approach allows as well the cases where a different space-time
evolution is taken into account. As a consequence of microscopic interactions, modified 
electric susceptibility, proper frequency and coupling constant  between electric field and 
polarization are postulated to occur, which are meant to reproduce a suitable 
behaviour of the refractive index perturbation. \\
In the actual construction of the model, we can refer to the aforementioned 
traveling perturbations, but, in order to be even much more general, we can allow a generic behavior for
the susceptibility $\chi (t,x,y,z)$, which is a scalar function of the space-time coordinates. Moreover, we allow an analogous 
nature also for $\omega_0 (t,x,y,z)$, the latter being the proper frequency of the matter oscillators. 
As e.g. in \cite{barnett,suttorp-epl,suttorp-jpa}, we generalize the coupling between electric field and 
polarization field(s), in our case by introducing a scalar function $g (t,x,y,z)$ which plays the role of coupling.\\ 

The only reasonable constraint we impose are that asymptotically in time $\chi, \omega,g$  are 
constant, for obtaining well-defined particle states IN and OUT.  
Moreover, we can allow couplings with N oscillators, in such a way that we can manage with
N different susceptibilities, resonances and couplings.\\
There are two further features of our model which are very important: relativistic covariance
and {\color{black} quantization in a covariant gauge}. Covariance is fundamental for gettting rid of any ambiguity in actual
calculations, and in order to respect a fundamental requirement for a physical model. 
We stress that these are {\sl per s\`e} interesting contributions to the microscopically-grounded works 
on the subject of electromagnetic field in dielectric media, 
because a covariant generalization is presented and its canonical quantization {\color{black} in a 
covariant gauge} is performed.
Covariance, as is known, and is confirmed since the original work by Minkowski \cite{minkowski} 
and e.g. by \cite{post,Penfield-Haus}, 
is not simply a speculative exercise in the picture at hand, but allows to 
get the correct behavior of physical quantities when changing from a inertial observer to another one. 
An important example is represented by a uniformly travelling perturbation $v=$const moving along the $x$-direction,  where we can use
covariance for passing to the comoving frame, where the theory is static, and where
physical interpretation of the scattering process is much more perspicuous \cite{belcacciadalla-hawking}.\\ 
It is 
also worth mentioning a longstanding series of studies concerning electrodynamics of moving media, where covariance 
of the formalism plays a key-role. We limit 
ourselves to quote some seminal papers \cite{gordon,balasz,phammauquan,synge}, where phenomenological 
electrodynamics is adopted (ref. \cite{synge} is explicitly devoted to the dispersive case). In particular, 
\cite{gordon} and \cite{phammauquan} are also important references for the dielectric models in 
analogue gravity \cite{barcelo}. See also \cite{horsley}, where a covariant model for moving (homogeneous) 
dispersive dielectric media is studied and quantized in the Coulomb gauge.\\
As to quantization, we can provide a scheme of constrained quantization where
all subtleties of the theory are taken into account (see e.g. \cite{Gitman-Tyutin,Henneaux-Teitelboim}), 
and  still covariance plays an important
role for obtaining consistent quantization rules which would not be so clearly stated otherwise. As far as 
dispersive effects are not involving magnetic properties of the material, we can also consider that our 
model, for constant dielectric susceptibility, represents an improvement of the phenomenological model 
studied in \cite{Watson-Jauch}.\\
It is worth mentioning that  a very general and interesting picture is provided 
in \cite{suttorp-jpa}, 
where the susceptibility is a tensor field depending 
on space and time. Absorption is also included by means of a bath of oscillators whose 
interactions with the electromagnetic field originate dissipative effects. 
Still, as a novelty with respect to the aforementioned picture, we develop a formalism leaving room for   
covariance and also quantization in a covariant gauge, which are not treated therein.

\section{A covariant form for the Hopfield model}
\label{cov-model}

In the following, we take into consideration the electromagnetic field Lagrangian which is
apt for a dispersive lossless dielectric medium, as in Hopfield model \cite{hopfield,fano,kittel}. A further field,
representing material polarization, is introduced and coupled to the free electromagnetic field as
follows:
\beqna
\lagr_{em}:&=&\frac{1}{8\pi} \left( \frac{1}{c} \dot{\mathbf{A}} +\nabla \varphi\right)^2
-\frac{1}{8\pi} \left(\nabla\wedge \mathbf{A}\right)^2\cr
&+&\frac{1}{2\chi \omega_0^2} \left(\dot{\mathbf{P}}^2 -\omega_0^2 \mathbf{P}^2\right)
-\frac{g}{2 c} \left(\mathbf{P}\cdot \dot{\mathbf{A}}+\dot{\mathbf{A}}\cdot \mathbf{P}\right) \cr
&-&\frac{g}{2} \left(\mathbf{P}\cdot \nabla \varphi+\nabla \varphi\cdot \mathbf{P}\right).
\eeqna
As an example, a traveling perturbation is described by introducing
$\chi (x-v t,y,z), \omega_0 (x-v t,y,z), g(x-v t,y,z)$.\\
Only a first-principle introduction of the Kerr effect, like e.g. the one obtained by introducing
a fourth power of $\mathbf{P}$ would require substantial modifications (because of the non-linear term),
but we do not pursue this problem herein.\\
In order to pursue a more standard calculation for inferring particle creation, 
{\color{black} we introduce a covariant generalization of the Hopfield model in 3+1 dimensions:}
this would make easier to find out an inner product with respect to which one could
calculate the Bogoliubov coefficients in order to check if particle creation occurs
in the standard way. The problem consists, of course, in finding a covariant form
for the polarization part of the Hopfield Lagrangian. This is not a so trivial task.
The main problem is represented by the kinetic part of the
polarization field Lagrangian; with this aim, let us introduce $v^\mu$ as 
the 4-velocity of the bulk dielectric medium (we mean the velocity
of the dielectric sample, not the one of the dielectric perturbation). Then, the covariant lagrangian density is:
\beqnl
\lagr = &&-\frac{1}{16\pi} F_{\mu \nu} F^{\mu \nu}
-\frac{1}{2\chi\omega_0^2} \left[ (v^\rho \pa_\rho P_\mu) (v^\sigma \pa_\sigma P^\mu) \right]\cr
&&
+ \frac{1}{2\chi}  P_\mu P^\mu
-\frac g{2 c} (v_\mu P_\nu- v_\nu P_\mu) F^{\mu \nu}.
\eeqnl
{\color{black} Minkowski metric $\eta_{\mu\nu}$ is chosen with the standard signature for quantum field theory: 
$(+,-,-,-)$.} 
The latter model appears to
be the principal candidate in our consideration, because it provides a field equation for $P_\mu$ which, for
constant $\chi$, gives rise e.g. in the eikonal approximation to the correct covariant dispersion relation.\\
{\color{black} We have introduced a 4-vector in order to describe the polarization field 
in a covariant form. We have introduced, as a consequence, also the component $P_0$ 
of the field, which is absent in the rest frame. Actually, this new component is not an 
independent one, it depends on the 
spatial components, as can be easily ascertained by constructing the 4-vector corresponding to the 
standard polarization field, see e.g.
\cite{Penfield-Haus,DeGroot-Suttorp}. Indeed, 
the following condition has to be implemented:
\beq
v^\mu P_\mu =0,
\label{transverse}
\eeq
which is required at the level of the classical theory for the polarization vector. Note that in the 
rest frame, where $v^\mu=(c,0,0,0)$, the above condition amounts to $P_0=0$.} 
We assume this condition too. It is worth mentioning that this condition is the
correct one for our harmonic oscillator field $P$ coupled to the electromagnetic field, regardless
of its specific nature of polarization field.\\
As to the field equations, 
in the case of the electromagnetic field we obtain
\beqnl
-\frac{1}{4\pi} \partial_\nu F^{\nu \mu} -v^\nu \partial_\nu  \left(\frac{g}{c} P^\mu\right) +v^\mu \partial_\nu \left(\frac{g}{c} P^\nu\right)=0,
\label{electro-m}
\eeqnl 
and for the polarization field we have
\beqnl
-v^\alpha \partial_\alpha \left(\frac{1}{\chi \omega_0^2} \right) v^\beta \partial_\beta P^\nu-\frac{1}{\chi} P^\nu+\frac{g}{c} v_\rho F^{\rho \nu} =0.
\eeqnl
By contracting with $v_\mu$ the equation for the electromagnetic field  and taking into account (\ref{transverse})
we get $\partial_\mu (E^\mu+4\pi \frac{g}{c} P^\mu)=0$, where $E^\mu:=v_\nu F^{\nu\mu}$.
This is just the Gauss law for the electric induction field $D^\mu :=E^\mu+4\pi \frac{g}{c}  P^\mu$. This is the right condition
to identify $P^\mu$ as a polarization field and is required by compatibility among the transversality condition and the equations
of motion. It is also useful to define the induction tensor $G^{\mu\nu}$, which is such that $
D^\nu= v_\mu G^{\mu\nu},\ 
H_\mu= \frac 12 v^\nu \epsilon_{\mu\nu\rho\sigma} G^{\rho\sigma} :=\frac 12 v^\nu \epsilon_{\mu\nu\rho\sigma} F^{\rho\sigma}=B_\mu$. 
{\color{black} Notice that, in absence of free charges and currents, Gauss law and Ampere law are summarized by the equation $\partial_\mu G^{\mu \nu}=0$, 
which, contracted with $v_\nu$, amounts to $\partial_\mu (E^\mu+4\pi \frac{g}{c} P^\mu)=0$.}\\
{\color{black} It is also easy to realize that, by introducing the $\Phi_\mu := (A_\mu, P_\mu)$, whose first four components coincide with $A_\mu$ 
and the remaining four components coincide with $P_\mu$ \footnote{$\Phi_\mu$ is the direct sum of $A_\mu$ and $P_\mu$.}  we obtain a 
field theory which is quadratic in $\Phi_\mu$}.

\section{Conserved scalar product for the model}

We can now determine the conserved scalar product associated to the covariant Hopfield model.
The first step in order to determine the scalar product is to complexify the fields. The complexified lagrangian density becomes:
\begin{eqnarray}
\lagr_{em}^{cov} &=& -\frac{1}{16\pi} F^*_{\mu \nu} F^{\mu \nu} -\frac{1}{2\chi\omega_0^2} \left[ (v^\rho \pa_\rho P^*_\mu) (v^\sigma \pa_\sigma P^\mu) \right.\cr
 &-& \left.
\omega_0^2 P_\mu^\ast P^\mu \right] -\frac g{2 c} P^*_\mu v_\rho F^{\rho\mu}-\frac g{2 c} P^\mu v^\rho F^*_{\rho\mu}. \label{complex}
\end{eqnarray}
A symmetry of (\ref{complex}) is $ A_\mu \mapsto e^{i\phi} A_\mu,\  P_\mu \mapsto e^{i\phi} P_\mu,\  
A^*_\mu \mapsto e^{-i\phi} A^*_\mu,\ P^*_\mu \mapsto e^{-i\phi} P^*_\mu$, 
where $\phi$ is a constant phase. The associated conserved quantity can be computed by means of the usual Noether method. 
The computation is immediate and gives the conserved current:
\begin{eqnarray}
{\mathcal J}^\mu &=&\frac i2 \left[ \frac 1{4\pi} F^{*\mu\nu} A_\nu +\frac{1}{\chi \omega_0^2} v^\rho \partial_\rho P^{*\sigma} P_\sigma v^\mu \right. \cr
&-& \left. \frac g{c} (P^{*\mu} v^\rho -P^{*\rho} v^\mu) A_\rho -c.c.
\right].
\end{eqnarray}
Indeed, a direct computation shows that on the solutions of the equations of motion ${\mathcal J}^\mu$ satisfies 
$\partial_\mu {\mathcal J}^\mu =0$.
Thus, the standard argument shows that on any spacelike slice $\Sigma_t$ the quantity 
$Q:=\int_{\Sigma_t}{\mathcal J}^0 d^3x $ 
does not depend from $t$. This defines a conserved (Hermitian) quadratic form $Q$ on $\Phi_\mu=(A_\mu, P_\mu)$:
\begin{eqnarray}
Q((A_\mu, P_\mu))&=&\frac i2 \int_{\Sigma_t}
\left[ \frac 1{4\pi} F^{*0\nu} A_\nu +\frac{1}{\chi \omega_0^2} v^\rho \partial_\rho P^{*\sigma} P_\sigma v^0\right.\cr
&-&\left. \frac g{c} (P^{*0} v^\rho -P^{*\rho} v^0) A_\rho -c.c.
\right] d^3x.
\end{eqnarray}
This gives the conserved scalar product by means of the usual polarization formula. 
Then we obtain the conserved scalar product:
\begin{eqnarray}
&&\pmb((A_\mu, P_\mu)\pmb |(\tilde A_\mu, \tilde P_\mu )\pmb ) \cr 
&&=\frac i2 \int_{\Sigma_t}
\left[\frac 1{4\pi} F^{*0\nu} \tilde A_\nu +\frac{1}{\chi \omega_0^2} v^\rho \partial_\rho P^{*\sigma}
\tilde P_\sigma v^0 \right.\cr
&&\left. -\frac g{c} (P^{*0} v^\rho -P^{*\rho} v^0) \tilde A_\rho
 -\frac 1{4\pi} \tilde F^{0\nu} A^*_\nu\right.\cr
&&\left. -\frac{1}{\chi \omega_0^2} v^\rho \partial_\rho \tilde P^{\sigma} P^*_\sigma v^0
+\frac g{c} (\tilde P^{0} v^\rho -\tilde P^{\rho} v^0) A^*_\rho
\right]d^3x.
\label{scaprod}
\end{eqnarray}
This scalar product is very important in relation to the quantization of the model. Indeed, it allows
to define positive and negative norm states for the solutions of the field equations, i.e. it allows to
define in a proper way particles and antiparticles respectively. A proper quantization for the model
is discussed in the following section.\\ 
For example, if $\chi(t,\mathbf x)=\chi_0,\omega_0 (t,\mathbf x)=\omega_0, g(t,\mathbf x)=1$ and we work in the lab frame, 
the scalar product among plane waves (with on shell momenta)
\begin{eqnarray}
(A^\mu, P^\mu)&=&({\mathcal A}^\mu e^{-i\omega t+i \mathbf {k}\cdot \mathbf x}, \cr
&-&i \frac {\chi_0}{\omega_0^2-{\omega}^2} \frac {\omega}{c} {\mathcal A}^{\mu} e^{-i\omega t+i \mathbf {k}\cdot \mathbf x}),
\label{plane}
\end{eqnarray}
and
\begin{eqnarray}
(\tilde A^\mu, \tilde P^\mu)&=&(\tilde{\mathcal A}^\mu e^{-i\tilde \omega t+i \mathbf {\tilde k}\cdot \mathbf x}, \cr
&-&i \frac {\chi_0}{\omega_0^2-{\tilde \omega}^2} \frac {\tilde\omega}{c} \tilde{\mathcal A}^{\mu} e^{-i\tilde \omega t+i \mathbf {\tilde k}\cdot \mathbf x}),
\label{plane-tilde}
\end{eqnarray}
{\color{black} where ${\mathcal A}^{\mu},\tilde{\mathcal A}^\mu $ in (\ref{plane}) and in (\ref{plane-tilde}) stand for constant 4-vectors representing the amplitude of the respective plane waves}, 
is
\begin{eqnarray}
&&\pmb((A_\mu, P_\mu) \pmb| (\tilde A_\mu, \tilde P_\mu) \pmb) \cr
&=&
\frac \omega{c} \left[ \frac 1{4\pi} + \frac {\chi_0 \omega_0^2}{(\omega_0^2-\omega^2)^2} \right] \mathbf A^* \cdot \tilde {\mathbf A}\ \delta^{(3)} (\mathbf k-\mathbf {\tilde k}),
\label{esscal}
\end{eqnarray}
where ${A}^\mu=(0, \mathbf A)$ and $\tilde{A}^\mu=(0, \mathbf {\tilde A})$.

\section{Quantization of the covariant model}

The introduction of the condition (\ref{transverse}) amounts to a constraint to be imposed on the system.
This affects also the quantization of the covariant Hopfield model, in the sense that the covariant form of
the Heisenberg commutation relations has to be consistent with the constraints of the theory.\\
As to the electromagnetic part of the Lagrangian, the procedure we follow slightly departs from what 
could be considered as 
standard, e.g. the usual
quantization under covariant gauge conditions like the Lorentz gauge $\partial_\mu A^\mu =0$, and where  
a Gupta-Bleuler formalism (see e.g. \cite{zuber})
can be adopted in order to get rid of spurious degrees of freedom, leaving only transverse (physical)
ones. Instead, we follow the Dirac approach in which all first-class constraints which are associated 
with the gauge freedom are first reduced to second-class ones by means of suitable explicit gauge-fixing 
terms in the Lagrange formalism, and then quantized \cite{dirac-book,dirac-papers}. Constraints are then implemented operatorially, 
rather than in a weak sense (as in the Gupta-Bleuler approach). See e.g. 
\cite{Gitman-Tyutin,Henneaux-Teitelboim,Rothe,barcelos-neto}.\\ 
In order to perform a complete quantization of the full Hopfield covariant model, in a
covariant gauge, we have to take into account both the electromagnetic part and the polarization
part of the Lagrangian. Both these parts require a suitable implementation of the constraints. 
We can add the constraints to the covariant Lagrangian, thus obtaining:
\begin{eqnarray}
\lagr_c : =&& -\frac{1}{16\pi} F_{\mu \nu} F^{\mu \nu}
-\frac{1}{2\chi \omega_0^2} \left[ (v^\rho \pa_\rho P_\mu) (v^\sigma \pa_\sigma P^\mu) \right] \cr
&&+ \frac{1}{2\chi}  P_\mu P^\mu
-\frac g{2 c} (v_\mu P_\nu- v_\nu P_\mu) F^{\mu \nu} \cr
&&+B (\pa_\mu A^\mu) + \frac{\xi}{2} B^2
+\lambda (v_\mu P^\mu),
\end{eqnarray}
where $B$ plays the usual role of auxiliary hermitian scalar field, also known as $B$-field
\cite{Nakanishi-Ojima,Gitman-Tyutin}, and $\xi$ is a constant which is useful for reproducing
various gauge conditions (the so-called $R_\xi$-gauges). {\color{black} The equations of motions are 
\beqnl
-\frac{1}{4\pi} \partial_\nu F^{\nu \mu} -v^\nu \partial_\nu  \left(\frac{g}{c} P^\mu\right) +v^\mu \partial_\nu \left(\frac{g}{c} P^\nu\right)\cr
+\partial^\mu B 
=0, \label{eq-ele-f}\\
-v^\alpha \partial_\alpha \left(\frac{1}{\chi \omega_0^2} \right) v^\beta \partial_\beta P^\nu-\frac{1}{\chi} P^\nu+\frac{g}{c} v_\rho F^{\rho \nu}\cr
+\lambda v^\mu =0, \label{eq-p-f}\\
\partial_\nu A^\nu+\xi B=0. \label{eq-b-f}
\eeqnl
Furthermore, as far as the conjugate momenta are concerned, we obtain:
\beqnl
\frac{\partial \lagr_c}{\partial \partial_t A_0} &=& :\Pi_A^0=\frac{B}{c},\\
\frac{\partial \lagr_c}{\partial \partial_t A_i} &=&: \Pi_A^i=- \frac{1}{4\pi c} (\pa^0 A^i-\pa^i A^0) \cr 
&&{\hphantom{ = \Pi_A^i=} }            -\frac g{c^2} (v^0 P^i- v^i P^0),\\
\frac{\partial \lagr_c}{\partial \partial_t B} &=&: \pi_B= 0,\\
\frac{\partial \lagr_c}{\partial \partial_t P_\mu} &=&: \Pi_P^\mu=-\frac{1}{\chi \omega_0^2 c} v^0 v^\sigma
\partial_\sigma P^\mu ,\\
\frac{\partial \lagr_c}{\partial \partial_t \lambda} &=&: \pi_\lambda= 0.
\eeqnl
As to the classical Hamiltonian density, according to the standard procedure we have:
\beqnl \label{hamilt}
\ham &=& (\pa_t A^\mu) \Pi_{A\; \mu}+
(\partial_t P^\mu) \Pi_{P\; \mu} - \lagr_c + u \pi_\lambda \cr
&+&y \pi_B+ z (\Pi_A^0-\frac Bc)\cr
&=&
2\pi c^2 (\Pi_A^i)^2 + \frac{1}{16\pi} F_{ij} F^{ij} +c  A_0 (\partial_i \Pi_{A\; i}) \cr
&+& {4\pi g} (v_0 P_i-v_i P_0)   \Pi_{A\; i} - c \frac{v^k}{v_0} (\pa_k P^\mu) \Pi_{P\; \mu} \cr 
&-&\frac{\chi \omega_0^2 c^2}{2 (v^0)^2} \Pi_{P\; \mu} \Pi_P^\mu - \frac{1}{2\chi} P_\mu P^\mu
+\frac{2 \pi g^2}{c^2} (v_0 P_i-v_i P_0)^2\cr
 &+&
 \frac g{2 c} (v_i P_j- v_j P_i) F^{ij} -B (\pa_i A^i) -\frac{\xi}{2} B^2 \cr
&-&
\lambda (v_\mu P^\mu) +
u \pi_\lambda+y \pi_B 
+ z (\Pi_A^0-\frac{B}{c}).
\eeqnl
 In our model there are three primary constraints $\pi_B, \Pi_A^0-B/c,\pi_\lambda$. }
From the analysis of Poisson brackets of the primary constraints with the Hamiltonian
we can find the following complete set of constraints: 
\beqnl
\Gamma_1 &=& \pi_B ,\\
\Gamma_2 &=& \Pi_A^0-\frac{B}{c},\\
\Gamma_3 &=& v_\mu P^\mu,\\ 
\Gamma_4 &=& v_\mu \Pi_P^\mu,\\ 
\Gamma_5 &=&  \lambda,\\ 
\Gamma_6 &=& \pi_\lambda. 
\eeqnl
We have taken into account that all functions of the constraints giving rise to the same submanifold $\Gamma_i=0$, $i=1,2,3,4,5,6$ are
to be considered equivalent, and this allows us to get the former simplified expressions for $\Gamma_4,\Gamma_5$. See also
\cite{Henneaux-Teitelboim,Gitman-Tyutin}. $\Gamma_1,\Gamma_2,\Gamma_6$ represent primary second-class constraints of the 
theory, and they appear explicitly in the expression of our constrained Hamiltonian defined above. {\color{black} We provide in the following 
some more details of the calculations. \\
In order to be more explicit, we get
\beqnl
\left\{ \pi_B, H\right\} &=& \partial_iA^i +\xi B+\frac{z}{c},\label{pib}\\
\{ \Pi^0_A-B, H\} &=&-c \partial_i \Pi_{Ai}-y,\label{pia0}\\
\{ \pi_\lambda, H\} &=&-v_\mu P^\mu,\label{pila}
\eeqnl
which represent the Poisson brackets with the primary second-class constraints. 
The first equation (\ref{pib}) fixes $z$ and is the same one obtains in QED \cite{Gitman-Tyutin}, and 
implies the same conditions. 
The second equation (\ref{pia0}) 
determines $y$. The third equation (\ref{pila}) introduces  a 
second-class second-stage constraint. Its Poisson bracket with the Hamiltonian is 
\beq
\{ v_\mu P^\mu, H\} =-\frac{\chi \omega_0^2 c^2}{(v^0)^2} v_\mu \Pi_P^\mu -c \frac{v^k}{v^0} \partial_k (v_\mu P^\mu) 
\label{vimupimu}
\eeq
which, on the sub manifold defined by the above constraints amounts to requiring 
$v_\mu \Pi_P^\mu =0$, which is then a second-class third-step constraint. Its Poisson bracket with 
the Hamiltonian is
\beq
\{ v_\mu \Pi_P^\mu, H\} =\frac{1}{\chi} v_\mu P^\mu +c \frac{v^k}{v^0} (\partial_k \delta)  (v_\mu \Pi_P^\mu)+v^\mu v_\mu \lambda, 
\label{vimupaimu}
\eeq
which, on the submanifold of the previous constraints implies $\lambda=0$. At this point we have a complete 
set of constraints, because 
\beq
\{ \lambda, H\} =u
\eeq
simply fixes $u=0$. See also below.}\\

We can define the matrix $\{C_{ij} \}$ whose entries are $C_{ij}:= \{ \Gamma_i,\Gamma_j \}$, 
where $\{,\}$ stay for the Poisson brackets. In particular, restoring the dependence on spacetime variables 
(time is fixed), with $x=(t,\mathbf{x})$, we get the following non-zero entries: 
$
\{ \Gamma_1 (t,\mathbf{x}),\Gamma_2 (t,\mathbf{y})\} = \delta^{(3)} (\mathbf{x}-\mathbf{y})/c,\ 
\{ \Gamma_3 (t,\mathbf{x}),\Gamma_4 (t,\mathbf{y})\} = v_\mu v^\mu  \delta^{(3)} (\mathbf{x}-\mathbf{y}),\ 
\{ \Gamma_5 (t,\mathbf{x}),\Gamma_6 (t,\mathbf{y})\} =  \delta^{(3)} (\mathbf{x}-\mathbf{y})$. 
Let us explicit the Dirac brackets as provided from the theory of constrained systems:
\beq
\{{\mathcal A},{\mathcal B}\}_D =\{{\mathcal A},{\mathcal B}\} - \{{\mathcal A},\Gamma_i \} C^{-1}_{ij} \{\Gamma_j,{\mathcal B}\}. 
\eeq
We recall that, in a less synthetic form, in the previous formula one has to take into account that, 
by introducing collective symbols for phase-space variables 
$
\{X_l\} =\{A^\mu,P^\mu,B,\lambda\},\ 
\{\bar{\Pi}_l\}=\{\Pi_A^\mu,\Pi_P^\mu,\pi_B,\pi_\lambda\}
$,   
we get:
\beqnl
&&\{{\mathcal A},{\mathcal B}\} =\int d^3 {z} \left(\frac{\delta {\mathcal A}}{\delta X_l (t,\mathbf{z})}
\frac{\delta {\mathcal B}}{\delta \bar{\Pi}_l (t,\mathbf{z})} \right. \cr 
&-& \left. \frac{\delta {\mathcal A}}{\delta  \bar{\Pi}_l (t,\mathbf{z})}
\frac{\delta {\mathcal B}}{\delta X_l (t,\mathbf{z})}\right),\cr
&&\{{\mathcal A},\Gamma_i \} C^{-1}_{ij} \{\Gamma_j,{\mathcal B}\} \cr
&=&
\int d^3 {u} d^3 {w} \{{\mathcal A},\Gamma_i (t,\mathbf{u})\} 
 C^{-1}_{ij} ( \mathbf{u},\mathbf{w}) \{\Gamma_j (t,\mathbf{w}),{\mathcal B}\};
\eeqnl
a summation convention on repeated indices is understood. One can also determine the Lagrange multipliers {\color{black} $u,y,z$ that appear 
for primary second-class constraints $\Gamma_1,\Gamma_2,\Gamma_6$ in the constrained Hamiltonian, as shown in our previous discussion.}
It is now easy to show that the following Dirac brackets hold true:
\beqnl
&\{ A^\mu (t,\mathbf{x}),\Pi_A^\nu (t,\mathbf{y})\}_D := 
\eta^{\mu \nu} \delta^{(3)} (\mathbf{x}-\mathbf{y}), \label{electro-vec}\\
&\{ P^\mu (t,\mathbf{x}),\Pi_P^\nu (t,\mathbf{y})\}_D \cr
&:= \left(\eta^{\mu \nu} - \frac{1}{v_\rho v^\rho} v^\mu v^\nu \right) \delta^{(3)} (\mathbf{x}-\mathbf{y}),\label{polvec}\\
&\{ B (t,\mathbf{x}), \pi_B (t,\mathbf{y})\}_D := 0,\label{par-b}\\
&\{ \lambda (t,\mathbf{x}), \pi_\lambda (t,\mathbf{y})\}_D :=0,
\eeqnl
which represent the basic ingredients for the quantization. Indeed, according to 
Dirac quantization scheme, we have to impose on quantum operators:  
\beq
[\hat{X}^l (t,\mathbf{x}), \hat{\bar{\Pi}}_k (t,\mathbf{y})]:=i\hbar  
 \{{X}^l (t,\mathbf{x}), {\bar{\Pi}}_k (t,\mathbf{y})\}_D.
\eeq
Moreover, constraints $\Gamma_i=0$, $i=1,\ldots,6$, are implemented as operators (cf. 
\cite{Gitman-Tyutin}, in particular p. 132): 
\beqnl
&&\hat{\pi}_B=0,\\
&&\hat{\Pi}_A^0=\frac{\hat{B}}{c},\\
&& v_\mu \hat{P}^\mu=0,\\ 
&& v_\mu \hat{\Pi}_P^\mu=0,\\ 
&& \hat{\lambda}=0,\label{lambda-cond}\\
&& \hat{\pi}_\lambda=0\label{pilambda-cond}.
\eeqnl 
{\color{black} As to the Hamiltonian, we have 
\beqnl \label{hamilt-field}
H &=& \int\; d^3 x \bigl[ 
2\pi c^2 (\Pi_A^i)^2 + \frac{1}{16\pi} F_{ij} F^{ij} +c  A_0 (\partial_i \Pi_{A\; i}) \cr
&+& {4\pi g} (v_0 P_i-v_i P_0)   \Pi_{A\; i} - c \frac{v^k}{v_0} (\pa_k P^\mu) \Pi_{P\; \mu} \cr 
&-&\frac{\chi \omega_0^2 c^2}{2 (v^0)^2} \Pi_{P\; \mu} \Pi_P^\mu - \frac{1}{2\chi} P_\mu P^\mu
+\frac{2 \pi g^2}{c^2} (v_0 P_i-v_i P_0)^2\cr
 &+&
 \frac g{2 c} (v_i P_j- v_j P_i) F^{ij} -B (\pa_i A^i) -\frac{\xi}{2} B^2 \cr
&-&
\lambda (v_\mu P^\mu) +
u \pi_\lambda+y \pi_B 
+ z (\Pi_A^0-\frac{B}{c})\bigr].
\eeqnl
As we are dealing with a model in which only second class primary constraints appear, we can also 
explicit the Lagrange multipliers as functions of the canonical variables. By taking into account our 
previous analysis, we can limit our attention only to $y,z$, which display the only non-trivial behaviour, 
and replace $y=-c \partial_i \Pi_{Ai}$ (cf. (\ref{pia0})), $z=-c (\partial_iA^i +\xi B)$ (cf. (\ref{pib})) in (\ref{hamilt-field}). (We recall that 
$u=0=\lambda$).}\\
It is interesting to note that, according to (\ref{lambda-cond}),(\ref{pilambda-cond}), $\lambda$ and its 
conjugate variable $\pi_\lambda$ are `expelled' by constrained quantization, in the sense that they are 
reduced both to the zero operator. We can provide the following interpretation.  {\color{black}
Condition (\ref{transverse}) holds true as a consequence of the antisymmetric character of the induction tensor $G^{\mu\nu}$ 
defined at the end of sec. \ref{cov-model}.} Indeed, we easily get $v_\nu D^\nu= v_\nu v_\mu G^{\mu\nu}=0$,
which, being $D^\mu :=E^\mu+4 \pi {\color{black}\frac{g}{c}} P^\mu$, and being $v_\mu E^\mu:=v_\mu v_\nu F^{\nu\mu}=0$, 
necessarily implies also $v_\mu P^\mu=0$. This 
condition is then preserved by quantization. This fact corroborates the previous results concerning the quantization 
of the model.

\subsection{Scalar product and constraints}

We notice that we have introduced in our setting constraints even in the Lagrangian approach. As 
a consequence, we have to take into account how the scalar product (\ref{scaprod}) is modified because of 
the constraints. After complexification of the constrained Lagrangian $\lagr_c$, we find that it appears only a further term 
in the scalar product: {\color{black} if $(A_\mu, P_\mu,B,\lambda)$ stays for the direct sum of $A_\mu, P_\mu,B,\lambda$, we get} 
\begin{eqnarray}
&&\pmb((A_\mu, P_\mu,B,\lambda) \pmb |(\tilde A_\mu, \tilde P_\mu, \tilde B, \tilde \lambda )\pmb )\cr
&&=  \frac i2 \int_{\Sigma_t}
\left[\frac 1{4\pi} F^{*0\nu} \tilde A_\nu +\frac{1}{\chi \omega_0^2} v^\rho \partial_\rho P^{*\sigma}
\tilde P_\sigma v^0 \right.\cr
&&\left. -\frac g{c} (P^{*0} v^\rho -P^{*\rho} v^0) \tilde A_\rho
 -\frac 1{4\pi} \tilde F^{0\nu} A^*_\nu\right.\cr
&&\left. -\frac{1}{\chi \omega_0^2} v^\rho \partial_\rho \tilde P^{\sigma} P^*_\sigma v^0
+\frac g{c} (\tilde P^{0} v^\rho -\tilde P^{\rho} v^0) A^*_\rho \right. \cr
&&\left. -(B^\ast \tilde A^0 - \tilde B {A^0}^\ast)
\right]d^3x.
\label{scaprodcon}
\end{eqnarray}
Notice that, assuming the same fields as in (\ref{esscal}), we still obtain the same result.\\
It is also important to point out that  the former scalar product can be found as follows. We introduce the 
phase-space vectors $(\{X^l\},\{\bar{\Pi}_l\})$, {\color{black} to be intended again as a direct sum}, by using 
the symplectic form: 
\beq
\Omega:=i \left[
\begin{array}{cc}
0 & 1_{10\times 10}\cr
-1_{10\times 10} & 0
\end{array}
\right],
\eeq
where $1_{10\times 10}$ stays for the identity matrix $10\times 10$.
Then we can define the following scalar product:
\begin{eqnarray}
&&\langle (\{X^l\},\{\bar{\Pi}_l\}),(\{\tilde{X}^l\},\{\tilde{\bar{\Pi}}_l\}) \rangle \cr
&&:=
\int (\{X^{\ast l},\bar{\Pi}^\ast_l\}) \cdot \Omega (\{\tilde{X}^l\},\{\tilde{\bar{\Pi}}_l\})  d^3 x,
\end{eqnarray}
where $\cdot$ stays for the usual Euclidean scalar product. Then, by taking into account the definitions for $\bar{\Pi}_l$ one gets the same 
result as in (\ref{scaprodcon}). {\color{black} This is the extension to a canonical quantization scheme 
and to 3+1 dimensions  of the results obtained for 1+1 dimensions in \cite{finazzi-carusotto-pra}.\\
{\color{black} It is also possible to introduce the operator $\tilde{H}$ such that the Hamiltonian in (\ref{hamilt-field}) can 
be written as follows (see e.g. \cite{dimock-qm}):
\beqnl
H &=& \frac{1}{2} ( (\{X^l\},\{\bar{\Pi}_l\}), \tilde{H} (\{X^l\},\{\bar{\Pi}_l\}))\cr
&=&\frac{1}{2} \int (\{X^{\ast l},\bar{\Pi}^\ast_l\}) \cdot \tilde{H} (\{\tilde{X}^l\},\{\tilde{\bar{\Pi}}_l\})  d^3 x
\eeqnl
which is such that, by defining $\Psi := (\{X^l\},\{\bar{\Pi}_l\})$, one obtains the Hamiltonian 
equations in the form 
\beq
\dot{\Psi}=-i \Omega (\nabla_\Psi H) =-i \Omega \tilde{H} \Psi,
\eeq
where $\nabla_\Psi=(\{\partial_{X^l}\},\{\partial_{\bar{\Pi}_l}\})$ It is also evident that, due to the trivialization 
of $\lambda$ and $\pi_\lambda$, we can restrict our considerations to a $9\times9$ dimensional phase space in a straightforward 
way.}

\subsection{Construction of physical states and physical operators}

In this subsection, we delve into the problem of constructing physical states 
and observables in our canonical approach in covariant gauges. As well-known, 
the price to be paid for covariance is the appearance of negative norm states, and 
also of zero-norm ones. As in QED, one has to face with the problem of non-physical 
states, which have to be eliminated from the physical spectrum. Also the Fano 
diagonalization procedure has to be revisited in view of the particular gauge conditions. 
{\color{black} In the standard approach to quantum electrodynamics, the calculation 
of the S-matrix requires several ingredients. Lacking a clear and well-defined way to 
implement a full non-perturbative quantum field theory in the Heisenberg 
representation, a perturbative approach is implemented in the interaction representation, 
where the fields evolve as free fields (i.e. their evolution is prescribed by the 
free field Hamiltonian $H_0$). A delicate interplay of the interacting (renormalizable) theory with asymptotic 
free fields occurs (cf. e.g. \cite{Gitman-Tyutin} for the case of QED). Asymptotic free fields (IN and OUT fields) 
satisfy canonical commutation relations and their creation and annichilation operators satisfy free-field  
commutation rules. In the present case, in order to simplify a bit our picture, we limit ourselves to the 
homogeneous dielectric medium, without losses and without dielectric perturbations e.g. induced by means 
of the Kerr effect. This is not a so crude limitation, as in region asymptotically far from the traveling 
dielectric perturbation, homogeneity can be assumed to be preserved. Within the given approximation, we 
assume a quantization strategy which goes parallel to the one of \cite{huttner-lett}, i.e. we 
quantize the fields as free, with the coupling $g=0$, and then perform a Fano diagonalization 
for the physical part of the Hamiltonian (with $g\not =0$). We expect that the procedure is consistent with the 
original commutation relations for the fields, as is in \cite{huttner-lett} (cf. also \cite{barnett}).\\
{\color{black} Furthermore, we wish to avoid any problem with the dipole ghost (see e.g. \cite{Nakanishi-Ojima}) 
so we impose the Feynman gauge $\xi=4\pi$ in what follows. See also the following section.} 
} 
\\
We start from the polarization field. It is a vector field with four components, so a priori 
it would involve four degrees of freedom. Still, the transversality condition 
$v^\mu P_\mu =0$ (\ref{transverse}) reduces to three the degrees of freedom which 
are actually available. This is also evident from (\ref{polvec}), which in the rest frame 
reduces identically to zero. Indeed, the set of eight fields $\{P_\mu, \partial_0 P_\mu\}$ 
represents a complete but not independent set of initial conditions, because (\ref{transverse}) 
allows to express $P_0, \partial_0 P_0$ as a function of the set $\{P_i, \partial_0 P_i\}$ which 
is both complete and independent. 
We can choose, in particular, a set of polarization vectors 
$e^\mu_\lambda,\ \lambda=1,2,3$, such that 
\beq
\sum_{\lambda=1}^3 e^\mu_\lambda e^\nu_\lambda = \eta^{\mu \nu}-\frac{v^\mu v^\nu}{v_\rho v^\rho},
\label{pola-p}
\eeq
which can be easily implemented. {\color{black} See below.} 
As to $P^\mu$, we have
\beq
P^\mu=\sum_{\lambda=1}^3 \int d^4 p \frac{1}{N_p} \delta (DR) \left[ e^\mu_\lambda  b_\lambda (\mathbf{p}) e^{i p x}+h.c.\right],
\label{polp}
\eeq
where $DR$ indicates the {\color{black} free field} dispersion relation, 
and $N_p$ is a suitable normalization factor. Our choice for the basis is 
$\{e^\mu_1,e^\mu_2,\frac{v^\mu}{c}-p^\mu \frac{c}{\omega}\}$, where $\omega=v^\mu p_\mu$ and 
\beq
e_\lambda^\mu = (0,\mathbf{e}_\lambda), \quad \quad \quad \lambda=1,2,
\eeq
in such a way that, for $\lambda=1,2$ it holds $\mathbf{e}_\lambda\cdot \mathbf{e}_\lambda'=\delta_{\lambda\lambda'}$, and 
\beqnl
v_\mu e_\lambda^\mu &=& 0, \label{v-transv}\\
p_\mu e_\lambda^\mu &=& 0 \label{k-transv},
\eeqnl
where $p^\mu$ is the wave-vector, as usual. The third polarization is orthogonal to $e^\mu_1,e^\mu_2$, and 
also to $v^\mu$. For the operators $b_\lambda (\mathbf{p}) $ the following canonical commutation relations (CCR) hold: 
\beq
\left[ b_\lambda (\mathbf{p}), b^\dagger_{\lambda'} (\mathbf{q}) \right] = \delta_{\lambda \lambda'} \delta^{(3)} (\mathbf{p}-\mathbf{q}).
\eeq

For the electromagnetic field and the auxiliary field $B$ quantization requires some more efforts.  
We {\color{black} can introduce} a further set of polarization vectors $\bar{e}^\mu_\lambda,\ \lambda=0,1,2,3$ and of 
operators $a_\lambda (\mathbf{p})$ such that 
{\color{black} 
\beq
A^\mu=\sum_{\lambda=0}^3 \int d^4 p \frac{1}{W_p} \delta (DR) \left[ \bar{e}^\mu_\lambda  a_\lambda (\mathbf{p}) e^{i p x}+h.c.\right],
\eeq
{\color{black} where $W_p$ is a suitable normalization. Usually, one imposes}
\beq
\sum_{\lambda=0}^3 \bar{e}^\mu_\lambda \bar{e}^\nu_\lambda = \eta^{\mu \nu}.
\label{pola-a}
\eeq
In particular, one may choose the set of polarization vectors $\{\frac{v^\mu}{c},e^\mu_1,e^\mu_2,e^\mu_3\}$, where 
$ e^\mu_\lambda,\ \lambda=1,2,3$ satisfy (\ref{pola-p}). Still, 
purposefully, in order 
to allow a more direct comparison with common literature and with \cite{Gitman-Tyutin}, we use the Gitman-Tyutin basis 
we costruct in the following. Furthermore, 
in this section, and only in 
this section, we choose $v^\mu = \gamma (c,0,0,v)$. 
We choose $e_\lambda^\mu,\ \lambda=1,2$ as in the case of the polarization field. The third polarization 
can be chosen e.g.  as follows (cf. \cite{Gitman-Tyutin}) in the case of the electromagnetic field: 
\beq
e_3^\mu = - i \frac{1}{|\mathbf{p}|} p^\mu, \quad \quad p^0=|\mathbf{p}|.
\eeq
{\color{black} It is also useful to choose
\beq
e_0^\mu=-i \frac{1}{2|\mathbf{p}|}  (|\mathbf{p}|,-\mathbf{p}).
\eeq
Our choices are the same as in \cite{Gitman-Tyutin}, apart from a slightly different normalization of both the vectors. We shall indicate the basis  
\beq
\{-i\frac{1}{2|\mathbf{p}|} (|\mathbf{p}|,-\mathbf{p}), e^\mu_1,e^\mu_2,- i \frac{1}{|\mathbf{p}|} p^\mu\}
\label{gt-basis}
\eeq 
as the Gitman-Tyutin basis. We stress that in this basis 
\beq
e_0^\mu e_0^\nu \eta_{\mu\nu}=0=e_3^\mu e_3^\nu \eta_{\mu\nu},
\eeq
and 
\beq
e_0^\mu e_3^ \nu\eta_{\mu\nu}=-1.
\eeq}
We also get, in the case of the auxiliary field $B$, 
\beq
B=\int d^4 p \frac{1}{S_p} \delta (DR) \left[ \beta (\mathbf{p}) e^{i p x}+h.c.\right],
\eeq
where $S_p$ is a suitable normalization. We remark that both $A^\mu$ and $B$ are considered as free fields, and then DR is a free field 
dispersion relation.} Furthermore, the equation of motion 
for $B$ (\ref{eq-b-f})
implies that we can reduce to four the overall degrees of freedom for the fields $A^\mu,B$, and then 
only four operators in the set $a_\lambda,\beta$, with $\lambda=0,1,2,3$ are independent. 
We choose, in analogy with the discussion in \cite{Gitman-Tyutin} (sec. 4.2 therein), 
\beq
a_0 (\mathbf{p}) = \beta (\mathbf{p}).
\eeq
{\color{black} Commutation relations are the same as in \cite{Gitman-Tyutin} , as we are using the 
same basis for the electromagnetic sector of our model. 
We find 
\beq
[\beta (\mathbf{p}),\beta^\dagger (\mathbf{q})]=0,
\eeq
and 
\beq
[a_3 (\mathbf{p}), \beta^\dagger (\mathbf{q})]=-\delta^{(3)} (\mathbf{p}-\mathbf{q}), 
\eeq
and also 
\beq
[a_\lambda (\mathbf{p}), \beta^\dagger (\mathbf{q})]=0, \quad \quad \quad \lambda=1,2. 
\eeq
Also, we find the following  commutation brackets for $a_\lambda (\mathbf{p})$, $\lambda=1,2,3$: 
\beq
\left[ a_\lambda (\mathbf{p}), a^\dagger_{\lambda'} (\mathbf{q}) \right] = \delta_{\lambda \lambda'} \delta^{(3)} (\mathbf{p}-\mathbf{q}).
\eeq
All other commutation relations are equal to zero. In particular, it holds 
\beq
\left[ a_3 (\mathbf{p}), a^\dagger_{3} (\mathbf{q}) \right] = 0,
\eeq
consistently with \cite{Gitman-Tyutin}. We point out that in the Gitman-Tyutin basis, and in the Feynman gauge, 
one may recover the above commutation relations by first recovering the functional relations between the creation-distruction 
operators and the fields and their conjugate momenta. Cf. eg. chapter 7 in \cite{greiner-quantization}, in particular subsection 
7.3.1 therein. We also recall that, a different choice of the basis with respect to the Gitman-Tyutin one, leads to different 
operators $\tilde{a}_\lambda, \tilde{a}_\lambda^\dagger$, which are linearly related (by means of a unitary transformation) to the ones of the Gitman-Tyutin basis. Moreover, 
the same commutation relations as above can be found also in a gauge with $\xi\not=4\pi$ \cite{Gitman-Tyutin}. } \\

{\color{black} It is important to consider the particular form of the interaction term between the polarization field and the electromagnetic 
field. This term is proportional to 
\beq
E^\mu P_\mu. 
\eeq
We have to take into account that $P^\mu$ is constrained by the transversality condition with respect to the velocity $v^\mu$
(\ref{transverse}). We can also introduce the projection operator 
\beq
{\mathcal{P}}^{\mu \nu} := \eta^{\mu \nu} -\frac{v^\mu v^\nu}{v^\rho v_\rho},
\label{v-pro}
\eeq
which is such that $v_\mu {\mathcal{P}}^{\mu \nu}=0$. It is evident that ${\mathcal{P}}^{\mu \nu} P_\nu=P^\mu$ and 
${\mathcal{P}}^{\mu \nu} E_\nu=E^\mu$, so that the interaction term is transverse with respect to $v^\mu$. 
Furthermore, we are considering isotropic dielectric media, where $\mathbf{P} \propto \mathbf{E}$. Then, we 
expect also that $P^\mu \propto E^\mu$. Moreover, from the equations of motion we have 
\beq
\partial_\mu D^\mu = \partial_\mu (E^\mu +4\pi \frac{g}{c} P^\mu) =0. 
\label{eq-gauss}
\eeq
This equation, in the homogeneous case (which represents the asymptotic limit in the comoving frame when 
a traveling perturbation is present), implies both $\partial_\mu E^\mu=0$ and $\partial_\mu P^\mu=0$, due to the aforementioned 
constitutive equation. As a consequence, by passing to the Fourier representation, this amounts to 
\beq
p_\mu E^\mu =0,
\eeq
i.e. also a transversality condition with respect to $p^\mu$ is implemented. If we introduce the further 
projection operator 
\beq
\bar{\mathcal{P}}^{\mu \nu} := \eta^{\mu \nu} -\frac{p^\mu p^\nu}{p^\rho p_\rho},
\label{p-pro}
\eeq
we have also 
\beq
E^\mu = \bar{\mathcal{P}}^{\mu \nu} E_\nu.
\eeq
As a consequence, in the interaction term we have 
\beq
E^\mu P^\nu \eta_{\mu \nu}=\bar{\mathcal{P}}^{\mu \rho} E_\rho  {\mathcal{P}}^{\nu \sigma} P_\sigma \eta_{\mu \nu},
\eeq
which means that the interaction term involves only polarizations which are transverse to both $v^\mu$ and $k^\mu$. 
In other terms, only physical transverse polarizations of the polarization field and of the electromagnetic field interact. 
Note that the projection operators ${\mathcal{P}}^{\mu \nu},\bar{\mathcal{P}}^{\mu \nu} $ commute. 
The scalar polarization $\lambda=0$, which involves only the electromagnetic field, and the longitudinal one $\lambda=3$, which 
involves both the electromagnetic field and the polarization field,  correspond to interaction-free parts 
of the field, which decouple from the physical spectrum, as shown below. In particular, we stress that the longitudinal 
component of the polarization field does not participate to any physical process (it corresponds to a free oscillator 
which has no interaction with the electromagnetic field and that cannot enter any asymptotic (physical) scattering state).}

We follow the discussion in \cite{Gitman-Tyutin} in order to construct states. {\color{black} As remarked above, in} our 
framework, we have very important simplifications to be taken into account, due to the fact that our model 
is free (in terms of path integral approach, it is Gaussian), so that we can avoid discussing Ward identities, 
and also the in-field formalism.\\ 

The space of states ${\mathcal{R}}$ is such that, given the vacuum state
\beq
\beta (\mathbf{p}) |0\rangle = a_\lambda (\mathbf{p}) |0\rangle=b_\lambda (\mathbf{p}) |0\rangle =0, \quad \quad \lambda=1,2,3,
\eeq
all states are spanned by states of the form 
\beq
(\beta^\dagger)^m  (a^\dagger_\lambda)^n (b^\dagger_\lambda)^l  |0\rangle, \quad \quad \lambda=1,2,3,
\eeq
where for simplicity of notation we have left implicit the dependence on momenta of the operators and where 
$l,m,n\in {\mathbb{N}}$. The space ${\mathcal{R}}$ contains states with zero norm and negative norm as well, 
due to the fact that {\color{black} $a_3, a_3^\dagger,\beta,\beta^\dagger$} satisfy {\sl non-canonical} commutation relations. We can easily 
confirm the presence of negative norm states as follows. Let us define {\color{black}
\beqnl
d_0 &:=& \frac{1}{\sqrt{2}} (a_3+\beta), \label{dzero}\\
d_3 &:=& \frac{1}{\sqrt{2}} (a_3-\beta). \label{dtre}
\eeqnl
Then we obtain the commutation relations}
\beq
[d_0 (\mathbf{p}), d_0^\dagger (\mathbf{q})]=-\delta^{(3)} (\mathbf{p}-\mathbf{q}), 
\eeq
and also 
\beq
[d_3 (\mathbf{p}), d_3^\dagger (\mathbf{q})]=\delta^{(3)} (\mathbf{p}-\mathbf{q}). 
\eeq
The state $d_0^\dagger |0\rangle$ has negative norm: 
\beq
\langle 0| d_0  (\mathbf{p}) d_0^\dagger  (\mathbf{p})|0\rangle = -\delta^{(3)} (0).
\eeq
 The space ${\mathcal{R}}$ is a space with indefinite metric. As remarked in \cite{Gitman-Tyutin}, 
this is the price to be paid in order to get an explicit Lorentz covariance. One could consider only 
the subspace ${\mathcal{R}}_\perp$ of vectors of the form 
\beq
(a^\dagger_\lambda)^n (b^\dagger_{\lambda'})^l  |0\rangle, \quad \quad \lambda=1,2,\quad  {\color{black} \lambda'=1,2,}
\eeq
which has a positive definite metric (it is an Hilbert space), but explicit 
covariance is lost. We also indicate the inverse formulas 
\beqnl
\beta &=&  \frac{1}{\sqrt{2}} (d_3-d_0),\\
a_3 &=&  \frac{1}{\sqrt{2}}  (d_3+d_0).
\eeqnl
It is useful to define the {\sl physical space}  ${\mathcal{R}}_{ph}$, which is spanned 
by vectors of the form 
\beq
(\beta^\dagger)^m  (a^\dagger_\lambda)^n (b^\dagger_{\lambda'})^l  |0\rangle, \quad \quad \lambda=1,2,{\color{black} \quad \lambda'=1,2}.
\label{phi-vect}
\eeq
It is a proper subspace of ${\mathcal{R}}$, because states generated by $a_3^\dagger$ {\color{black} and by $b_3^\dagger$} are absent. 
If we call ${\mathcal{R}}_0$ the subspace spanned by vectors (\ref{phi-vect}) with $m\not = 0$, 
we have 
\beq
{\mathcal{R}}_{ph}={\mathcal{R}}_\perp \oplus {\mathcal{R}}_0.
\eeq
States in ${\mathcal{R}}_0$ have zero norm and are orthogonal to any state in ${\mathcal{R}}_{ph}$. 
States in  ${\mathcal{R}}_{ph}$ can be identified also by means of the condition 
\beq
\beta |\Psi\rangle = 0,\quad \quad  |\Psi\rangle \in {\mathcal{R}}_{ph},
\eeq
as $\beta$ commutes with any other operator except for $a_3^\dagger$. This condition 
amounts to the usual Gupta-Bleuler condition, as it can be expressed as follow: 
\beq
\hat{B}^{(+)} (t,\mathbf{x}) |\Psi\rangle = 0,
\eeq
where $\hat{B}^{(+)}(t,\mathbf{x})$ is the positive frequency part of the operator  $\hat{B} (t,\mathbf{x})$ 
 (see \cite{Gitman-Tyutin} for details). 
The space ${\mathcal{R}}_{ph}$ can be described, as in   \cite{Gitman-Tyutin}, as a space 
whose non-zero norm vectors represent physical spaces, and vectors which differ for a zero-norm 
vector are physically equivalent:
\beq
|\Phi\rangle \simeq |\Psi\rangle \Longleftrightarrow (|\Phi\rangle - |\Psi\rangle)\in  {\mathcal{R}}_0.
\eeq

We take into account the construction of physical operators. In \cite{Gitman-Tyutin} it is 
shown that any physical operator $\hat{F}_{ph}$, i.e. any operator which is associated with a physical 
observable, should be such that 
\beq
\hat{F}_{ph} {\mathcal{R}}_{ph} \subset {\mathcal{R}}_{ph},
\eeq
i.e. ${\mathcal{R}}_{ph}$ should be invariant. In particular, the Hamiltonian $\hat{H}$ should be a physical 
operator, and indeed it is. Let us suppose that $|\Psi\rangle\in {\mathcal{R}}_{ph}$. Then we have 
\beqnl
\hat{B}^{(+)} (t,\mathbf{x}) \hat{H} |\Psi\rangle &=& [\hat{B}^{(+)} (t,\mathbf{x}),\hat{H}] |\Psi\rangle\cr
& =& i \hbar \partial_0 
\hat{B}^{(+)} (t,\mathbf{x}) |\Psi\rangle =0.
\eeqnl
This ensures that the space ${\mathcal{R}}_{ph}$ is left invariant under the action of the Hamiltonian $\hat{H}$.\\
{\color{black} We point out also that, for any observable $\hat{F}$ which does not depend explicitly on time, in the 
Heisenberg picture  it holds
\beq
i \hbar \frac{\partial \hat{F}}{\partial t}=[\hat{F},\hat{H}].
\eeq

{\color{black}
\subsection{The Hamiltonian and the Interaction Representation}

In the previous subsection, we have several times mentioned that we consider the 
fields as free. Implicitly, we have referred to the interaction representation, which is 
standard in quantum field theory as far as nontrivial interaction terms appear. 
There is no strict need to refer to this representation in our case, as the theory 
can be dealt with exactly. Still, there are nontrivial subtleties, which are related to 
the choice of the gauge. As is known, in standard QED the usual gauge for 
quantum electrodynamics is the so called Feynman gauge, i.e. $\xi=4\pi$. 
This gauge avoids to be faced with the appearance of the so-called ghost pole 
in the theory \cite{Nakanishi-Ojima}, which represents a nontrivial and quite hard problem to be 
dealt with. In the present case, the choice $\xi=4\pi$ is effective only in the 
interaction representation, in the sense that in the Heisenberg representation it 
is not available, as it is not difficult to realize. We defer the study in the 
latter representation to a future publication, as even in the homogeneous case 
(i.e. no traveling perturbation) there are many tricky formal problems which 
require extensive and long calculations. Herein, in the spirit of most books and 
studies in condensed matter physics, we adopt the interaction representation, 
so that field operators evolve freely, as dictated by the free Hamiltonian operator, 
and states evolve through the Dyson evolution operator which is constructed by 
means of the interaction term in the Hamiltonian. By referring to our case, we have 
\beq
H_0 = H_{em}+H_{pol},
\eeq
where $H_0$ is the free Hamiltonian contribution, with 
\beqnl \label{hamilt-qed}
H_{em} &=& \int\; d^3 x \bigl[ 
2\pi c^2 (\Pi_A^i)^2 + \frac{1}{16\pi} F_{ij} F^{ij} \cr
&+&c  A_0 (\partial_i \Pi_{A\; i}) -B (\pa_i A^i) -\frac{\xi}{2} B^2\bigr],
\eeqnl
and 
\beqnl \label{hamilt-pol}
H_{pol} &=& \int\; d^3 x \bigl[ 
-\frac{\chi \omega_0^2 c^2}{2 (v^0)^2} \Pi_{P\; \mu} \Pi_P^\mu - \frac{1}{2\chi} P_\mu P^\mu\cr
&-&c \frac{v^k}{v_0} (\pa_k P^\mu) \Pi_{P\; \mu}
+\frac{2 \pi g^2}{c^2} (M_{0i})^2
\bigr], 
\eeqnl
where we have defined 
\beq
M_{\alpha \beta}:=v_\alpha P_\beta-v_\beta P_\alpha.
\eeq
The interaction term is 
\beqnl \label{hamilt-int}
H_{int} &=& \int\; d^3 x \bigl[ 
\frac{g}{c} M_{0i} F^{0i} +\frac g{2 c} M_{ij} F^{ij}\bigr]\cr
 &=&\int\; d^3 x \bigl[ 
 \frac g{2 c} M_{\mu \nu} F^{\mu\nu}\bigr]=\int\; d^3 x \bigl[ 
 \frac g{ c} P_{\mu} E^{\mu}\bigr].
\eeqnl
One has to take into account that, in the interaction representation, $\Pi_A^i$ is considered at $g=0$, 
so a simplification in calculations occurs. Note that in (\ref{hamilt-pol}) the last term has been 
considered as a further contribution to the free Hamiltonian operator, despite the fact that 
it is proportional to $g^2$. This is due to the fact that this term `renormalizes' the proper frequency $\omega_0$ 
of the polarization field and, substantially, is a sort of improvement of the standard free Hamiltonian. See e.g. 
\cite{hopfield,kittel,huttner-lett}.\\
Furthermore, the Hamiltonian $H_{em} $ can be simplified by using the equations of motion for the electromagnetic field: 
indeed, the last three terms on the solutions of the equations of motion become 
\beqnl
&&c  A_0 (\partial_i \Pi_{A\; i}) -B (\pa_i A^i) -\frac{\xi}{2} B^2\cr
&& = B\partial_0 A_0-A_0 \partial_0 B +\frac{\xi^2}{2} B^2,
\eeqnl
and this is true both in the interaction representation with $g=0$ and in full interacting case, as a direct inspection 
confirms. Then we obtain 
\beqnl \label{hamilt-qed-os}
H_{em} &=& \int\; d^3 x \bigl[ 
2\pi c^2 (\Pi_A^i)^2 + \frac{1}{16\pi} F_{ij} F^{ij} \cr
&+&B\partial_0 A_0-A_0 \partial_0 B +\frac{\xi^2}{2} B^2\bigr],
\eeqnl
where one can immediately realize that the last three terms in the free electromagnetic 
Hamiltonian do contribute only to the unphysical polarizations, and then are 
substantially irrelevant for the physics at hand (their role is only formal, being a consequence of the 
covariance requirement, as seen).\\
We display the explicit expression of the Hamiltonian in terms of creation and annihilation operators. 
We recall our choices for the bases: the Gitman-Tyutin basis (\ref{gt-basis}) for the 
electromagnetic field, and 
$\{e^\mu_1,e^\mu_2,\frac{v^\mu}{c}-p^\mu \frac{c}{\omega}\}$ for the polarization field.  
We first fix the normalizations for the fields:
\beq
P^\mu=\sum_{\lambda=1}^3 \int \frac{d^3 p}{(2\pi)^{3/2}} \sqrt{\frac{\chi \omega_0^2 }{2\Omega_0}} 
\left[e^\mu_\lambda  b_\lambda (\mathbf{p}) e^{-i p x}+h.c.\right],
\label{pol-free}
\eeq
where the renormalized frequency is 
\beq
\Omega=\omega_0 \sqrt{1+4\pi g^2 \chi},
\eeq
and, for the electromagnetic field and the auxiliary field we get
\beq
A^\mu=\sum_{\lambda=0}^3 \int \frac{d^3 p}{(2\pi)^{3/2}} \sqrt{\frac{(2\pi)}{p_0}}  \left[ {e}^\mu_\lambda a_\lambda (\mathbf{p}) e^{-i p x}+
h.c.\right] 
\eeq
\beq
B=\frac{1}{4\pi} \int \frac{d^3 p}{(2\pi)^{3/2}} \sqrt{\frac{(2\pi)}{p_0}} p_0 \left[ \beta (\mathbf{p}) e^{-i p x}+h.c.\right].
\eeq
In the following, each contribution to the Hamiltonian is distinguished between transversal and non-trasversal. 
We know that $\hat{H}$ is at most quadratic in the creation and annihilation operators which span the 
space ${\mathcal{R}}$. Moreover, it can be written as follows:
\beq
\hat{H}=\hat{H}_{ph}+\hat{H}',
\eeq
where $\hat{H}_{ph}$ corresponds to the physical part of the Hamiltonian, which is the same as in the 
Coulomb gauge (cf. \cite{Gitman-Tyutin} for QED), and involves only transverse polarizations. $\hat{H}'$ 
is the remaining part, which involves unphysical polarizations, which are decoupled from the physical 
spectrum.\\
We develop our calculations for the specific case $v^\mu=(c,0,0,0)$ for simplicity; we get:
\beqnl
\hat{H}_{em, ph}&=& \int d^3 p\; p_0 \sum_{\lambda=1}^2 a^\dagger_\lambda (\mathbf{p},t) a_\lambda (\mathbf{p},t),\cr
\hat{H}_{pol, ph}&=& \int d^3 p\; \Omega_0 \sum_{\lambda=1}^2 b^\dagger_\lambda (\mathbf{p},t) b_\lambda (\mathbf{p},t),
\eeqnl
for the free part, and
\beqnl
\hat{H}_{int, ph}&=& i \frac{g}{c} \int d^3 p\; \sqrt{\frac{\pi \chi \omega_0^2}{p_0 \Omega}} c p_0 
\sum_{\lambda=1}^2\cr
&&\left[ \left(a_\lambda (\mathbf{p},t)  b_\lambda (-\mathbf{p},t)-
 a^\dagger_\lambda (\mathbf{p},t)  b^\dagger_\lambda (-\mathbf{p},t)\right)+\right.\cr
&&\left.  \left(a_\lambda (\mathbf{p},t)  b^\dagger_\lambda (\mathbf{p},t)-
 a^\dagger_\lambda (\mathbf{p},t)  b_\lambda (\mathbf{p},t)\right)\right].
\eeqnl
for the interaction term. We recall that the operators $a_\lambda (\mathbf{p},t)$ in the interaction representation are related to the 
ones $a_\lambda (\mathbf{p})$ in the Schroedinger and the ones $a_\lambda^h (\mathbf{p},t)$ in the Heisenberg representations as follows: 
\beqnl
a_\lambda (\mathbf{p},t) &=& e^{i\hat{H}_0 t} a_\lambda (\mathbf{p}) e^{-i\hat{H}_0 t},\\
a_\lambda^h (\mathbf{p},t) &=& e^{i\hat{H} t} a_\lambda (\mathbf{p}) e^{-i\hat{H} t}.
\eeqnl
Analogous formulas hold for $\beta (\mathbf{p},t),a_3 (\mathbf{p},t), b_\lambda (\mathbf{p},t)$. 
The contributions to the non-transversal (unphysical) part of the Hamiltonian are (we omit to use explicit arguments when
no ambiguity occurs)
\beqnl
\hat{H}'_{em}&=& \int d^3 p\; \biggr[\left(-\frac{p_0}{4} \right) \beta^\dagger \beta\biggr]
\eeqnl
for the electromagnetic field, and 
\beq
\hat{H}'_{pol}=\int d^3 p\; \Omega_0 b^\dagger_3 b_3
\eeq
for the polarization part. For the interaction term we get
\beqnl
&&\hat{H}'_{int}= \frac{g}{c} \int d^3 p\; \sqrt{\frac{\pi \chi \omega_0^2}{p_0 \Omega}}\cr 
&& \biggl( 
\frac{1}{2} c p_0 \left[ \left(\beta (\mathbf{p},t)  b_3 (-\mathbf{p},t)+
 \beta^\dagger (\mathbf{p},t)  b^\dagger_3 (-\mathbf{p},t)\right)+\right.\cr
&&\left.  \left(\beta (\mathbf{p},t)  b^\dagger_3 (\mathbf{p},t)+
 \beta^\dagger (\mathbf{p},t)  b_3 (\mathbf{p},t)\right)\right]\biggr).
\eeqnl
It is important to underline that matrix elements between physical states  of the $H'$ operators vanish:
\beq
\langle \Phi | H' |\Psi\rangle = 0,\quad \quad |\Phi\rangle, |\Psi\rangle \in {\mathcal{R}}_{ph}.
\eeq
Indeed, the term in $b_3^\dagger b_3$ vanishes because physical states don't contain any 
longitudinal quanta of polarization by construction. The other terms vanish as well. It may be interesting 
to point out that, in our construction, the condition 
\beq
(\partial_\mu P^\mu)^{(+)} |\Psi\rangle = 0
\eeq
is implemented for any $|\Psi\rangle \in {\mathcal{R}}_{ph}$. This condition can be understood as a 
transversality condition in a weak sense (as only for $\lambda=1,2$ the polarization field 
implements the transversality condition automatically, with $p_\mu e^\mu_\lambda=0$. 
Cf. the discussion which follows 
eq. (\ref{eq-gauss})).}\\

As to the physical part, we can find a linear canonical transformation (a Bogoliubov transformation, 
known as Fano transformation for this specific model \cite{hopfield,fano}), 
which  carries  $\hat{H}_{ph}$ into a simplified form 
involving quasi-particle states $\alpha_\lambda (\mathbf{p})$, with $\lambda=1,2$. 
In particular, as in the original Hopfield paper \cite{hopfield}, we impose for $\lambda=1,2$
\beq
[\alpha_\lambda (\mathbf{p}), \hat{H}_{ph}]=\omega (\mathbf{p}) \alpha_\lambda (\mathbf{p}),
\label{diago}
\eeq
and the corresponding commutation relations are
\beq
\left[ \alpha_i (\mathbf{p}), \alpha_j^\dagger (\mathbf{q}) \right] = \delta_{ij} \delta^{(3)} (\mathbf{p}-\mathbf{q}).
\eeq
In this way we can recover the same dispersion relation as for the full model, as it is easy to 
verify. Cf. also \cite{huttner-lett}. 

\subsection{A more general setting}

In concluding this section, we point out that the above model can be easily extended to the case of $N>1$
material harmonic oscillators coupled with the electromagnetic field. The simple substitutions $P^\mu
\mapsto P^\mu_{(k)}$, with $k=1,\ldots,N$, $\omega_0 \mapsto \omega_{0 (k)}$,
$\chi \mapsto \chi_{(k)}$ and $g\mapsto g_{(k)}$, lead to the desired form of the
Lagrangian:
\beqnl
\lagr_c : &=& -\frac{1}{16\pi} F_{\mu \nu} F^{\mu \nu} \cr
&-&\sum_{k=1}^N \left[\frac{1}{2\chi_{(k)}\omega_{0 (k)}^2} \left[ (v^\rho \pa_\rho P_{(k)\mu}) (v^\sigma \pa_\sigma P^\mu_{(k)}) \right]\right.\cr
&-&\left. \frac{1}{2\chi_{(k)}}  P_{(k) \mu} P^\mu_{(k)}
+\frac{g_{(k)}}{2 c} (v_\mu P_{(k) \nu}- v_\nu P_{(k) \mu}) F^{\mu \nu}\right]\cr
&+& B (\pa_\mu A^\mu) + \frac{\xi}{2} B^2
+\sum_{k=1}^N \lambda_{(k)} (v_\mu P^\mu_{(k)}).
\eeqnl
At the level of the constraints, in place of $\Gamma_3,\Gamma_4$ we get 2N constraints
$\Gamma_{3 (k)},\Gamma_{4 (k)}$, and, analogously, 2N constraints
$\Gamma_{5 (k)},\Gamma_{6 (k)}$. Fields $P^\mu_{(k)}$ satisfy:
\begin{eqnarray}
&\{ P^\mu_{(k)},\Pi_{P_{(l)}}^\nu \}_D \cr
&:= \delta_{(k) (l)}
\left(\eta^{\mu \nu} - \frac{1}{v_\rho v^\rho} v^\mu v^\nu \right) \delta^{(3)} (\mathbf{x}-\mathbf{y}).
\end{eqnarray}

\section{Example: model with a $v=$const traveling dielectric perturbation}

Let us consider the case where the dielectric perturbation induced by means of the Kerr effect is traveling 
with constant velocity $v$ in the lab frame. We can also assume that dependence on 
transverse coordinates $y,z$ is absent (which means that our dielectric perturbation is actually modelized as a 
dielectric slab infinitely extended in transverse directions). The relevance of a covariant approach is 
easily appreciated by taking into account that, in the reference frame comoving with the perturbation, 
one gets a static dependence of the parameters $\chi,\omega_0, g$ on $x'/\gamma$, where 
$x' = \gamma (x -v t),\ t' =  \gamma (t -\frac{v}{c^2} x)$ 
represent the Lorentz boost connecting lab and comoving frame. As a consequence of this, it is easily 
understood that energy is conserved in the comoving frame, i.e. it is possible to perform a variable separation 
involving the time coordinate in such a way that the energy $\omega'$ in the comoving frame is conserved. 
This result is very important and helpful in interpreting scattering in presence of the perturbation, and amply 
corroborates the relevance of a covariant approach and a consistent quantization. In particular, it is possible 
to quantize the system and to find out a scattering basis for the quantum fields in a 
straightforward way, without any problem arising because of a possible time-dependence of the perturbation. 
Moreover, covariance, together with a correct quantization, allows to find out a conserved inner product 
whose associated norm is fundamental in defining particles and antiparticles for the given model. Note that 
the norm sign is independent on the frame chosen, and then is an invariant concept (at it should be).\\ 
A further paper is 
dedicated to results and analysis about this specific model, with special reference to the 
question of analogous Hawking radiation in dielectric media \cite{belcacciadalla-hawking}.  We limit ourselves 
to point out that, as far as the full model with the uniformly travelling perturbation is concerned, 
in the lab frame one can still implement quantization by means of standard canonical commutation
relations. Indeed, the electromagnetic part can be quantized in the Coulomb gauge, without any change with respect to
the standard strategy (se e.g. \cite{hopfield}, or even \cite{barnett,suttorp-epl} for a more involved model). 
The polarization field part does not require a particular care in the definition of the variable conjugate to $\mathbf{P}$,
which is obtained by standard tools of Lagrangian formalism. One is lead to the following equal time commutation relations:
\beq
\left[P_i (t,\mathbf{x}), \frac{1}{\chi \omega_0^2 (t,\mathbf{y})} \partial_t P_j (t,\mathbf{y}) \right]= i \hbar \delta_{ij}
\delta^{(3)} (\mathbf{x}-\mathbf{y}).
\eeq
It can be easily shown that the conjugate momentum $\frac{1}{\chi \omega_0^2} \partial_t P_i$ leads to correct Hamiltonian equations
for the polarization field.\\
We stress again that, in the lab, the presence of the travelling perturbation induces an explicit dependence on $t$
of the total Hamiltonian of the model. This implies that energy is not conserved in this frame, still one can expect that
some sort of conservation occurs (cf. e.g. what happens in the case of the generalized Manley-Rowe identities \cite{belcacciadalla-hawking}).
This makes 
the quantization in the lab less clear and more problematic than the one in the comoving frame.

\section{Asymptotic behavior of solutions in the comoving frame for $v=$const}

The Hamiltonian (\ref{hamilt}) allows variables separation for the solutions. Defining:
\begin{eqnarray}
A^\mu(x,y,z,t)&=e^{-i\omega t+ik_yy+ik_zz}a^\mu(x), \cr
P^\mu(x,y,z,t)&=e^{-i\omega t+ik_yy+ik_zz}p^\mu(x), \cr
B(x,y,z,t)&=e^{-i\omega t+ik_yy+ik_zz}b(x), \cr
\lambda(x,y,z,t)&=e^{-i\omega t+ik_yy+ik_zz}l(x)
\end{eqnarray}
we obtain a second order system of ordinary differential equations for the variables $a^\mu(x)$, $p^\mu(x)$, $b(x)$ and $l(x)$. 
Since the equation for $l(x)$ is $l(x)=0$ we can omit this variable from the system.
The equation involving $b(x)$ is algebraic and can be used to simplify the other eight.
To this second order system we can associate a first order one by introducing:
\begin{eqnarray}
&&\alpha^\mu(x):=\partial_xa^\mu(x), \quad \pi^\mu(x):=\partial_xp^\mu(x), \cr
&& \beta(x):=\partial_x b(x),
\end{eqnarray}
then if $W(x):=(a^\mu(x),\alpha^\mu(x),p^\mu(x),\pi^\mu(x))$, we obtain the following system:
\beq
W'(x)=K_{16}W(x),
\eeq
where $K_{16}$ is a suitable $16\times 16$ operator. This can be written as $K_{16} = {\mathcal C}+{\mathcal R}(x)$, where ${\mathcal C}$ is a constant $16\times 16$ matrix and ${\mathcal R}(x)$ contains the non constant part. For simplicity, let us consider the case where only dielectric 
susceptibility varies; then ${\mathcal R}(x)$ has the form:
\begin{eqnarray}
{\mathcal R}(x) 
:=\left(
\begin{array}{cccc}
0_4 & 0_4 & 0_4 & 0_4 \\
 A_4 & B_4 & C_4 & D_4 \\
 0_4 & 0_4 & 0_4 & 0_4 \\
 0_4 & 0_4 & 0_4 & 0_4 \\
\end{array}
\right)
,
\end{eqnarray}
where $0_4$ is the $4\times 4$ identity matrix, and 
$A_4=-\frac{i v^0 \omega \omega_0^2 \chi (x)}{c (v^1)^2}I_4$, $B_4=\frac{\omega_0^2 \chi(x)}{c v^1}I_4$, 
$C_4=-\frac{i v^0 \omega \chi'(x)}{v^1 \chi (x)}I_4$, $D_4=\frac{\chi '(x)}{\chi (x)}I_4$, with $I_4$ the 
$4\times 4$ identity matrix. 
Under the hypothesis:
\beq
\int_a^\infty dx |{\mathcal R}(x)|<\infty,
\eeq
which physically can match very well the nature of travelling perturbation of $\delta \chi$ (see 
e.g. the theory displayed in \cite{eastham}), we can infer that, both as $x\to \infty$ and as $x\to -\infty$, 
the asymptotic behavior of solutions is governed by the eigenvalues of ${\mathcal C}$, which implies that 
the basis for $\delta \chi=0$ is asymptotically a good scattering basis also for the perturbed problem.  
To be more precise: the asymptotic region solutions are a scattering basis, and, moreover, solutions of the 
full equations asymptotically behave as the asymptotic region solutions, which then represent a good 
scattering basis.

\section{Conclusions}

We have presented a model aimed to a semi-phenomenological description of quantum electrodynamics
in presence of a dielectric perturbation in a dielectric medium. The standard Hopfield model has been
made fully covariant, and its quantization procedure has been discussed in detail. We stress that
the requirement of covariance is fundamental in order to allow a proper interpretation of measurable
quantities (e.g. quantum probabilities) as viewed by different (inertial) observers. E.g., in the
discussion of the analogue Hawking effect in dielectrics, a very important conceptual tool consists
in the analysis one performs in the comoving frame of the dielectric perturbation induced by the
Kerr effect. This request for covariance reflects itself in a more tricky quantization procedure,
which involves both the electromagnetic field, which represents a constrained system, as any gauge
theory, and, as such, requires a special quantization procedure, and also the polarization field,
due to a further constraint it carries into the lagrangian of the system. We have dealt the problem
both in a non-covariant approach and in a covariant one. {\color{black} We have also discussed how to 
identify the asymptotic behaviour of the solutions.\\
Our direct developments of the present work include: 1) a perturbative approach \cite{cacciatori-perturb}
for the model, 2) an application of the aforementioned approach} to the problem of photon pair creation by a helicoidal
rotating dielectric perturbation \cite{dallapiazza-letter}, 3) the nonperturbative study of the analogue
Hawking effect \cite{belcacciadalla-hawking},
together with the elaboration of a simplified model reproducing the basic features
of the full Hopfield model discussed here. 4) The description, at the perturbative level, of 
cosmological analogue situations \cite{westerberg}.

\section*{References}


\begin{thebibliography}{99}

\bibitem{heisenberg}
\newblock{W.~Heisenberg and H.~Euler, {\sl Zeitschr. Phys.} {\bf 98}, 714 (1936). 
English translation in arXiv:physics/0605038.}


\bibitem{schwinger}
J. Schwinger, Phys. Rev. 82, 664 (1951).

\bibitem{schwinger-sono}
J. Schwinger, Proc. Natl. Acad. Sci. 89, 4091–4093
(1992); 89, 11 118–11 120 (1992); 90, 958–959 (1993);
90, 2105–2106 (1993); 90, 4505–4507 (1993); 90, 7285–
7287 (1993); 91, 6473–6475 (1994).

\bibitem{luks}
A.~Luks, V.~Perinov\'a, Quantum Aspects of Light Propagation. Springer, Berlin (2009). 




\bibitem{hopfield}
\newblock{J.J.Hopfield, {\sl Phys. Rev.} {\bf 112}, 1555 (1958).}

\bibitem{fano}
\newblock{U.Fano, {\sl Phys. Rev.} {\bf 103}, 1202 (1956).}

\bibitem{kittel}
\newblock{C.Kittel, {\sl Quantum theory of solids.} Wiley, New York (1987).}


\bibitem{davydov}
A.S.Davydov, Teoria del solido. Mir, Moscow (1984). 

\bibitem{barnett}
B.Huttner and S.M.Barnett, Phys. Rev. A {\bf 46},  4306 (1992).


\bibitem{suttorp-epl}
L.G.~Suttorp and A.J.~van Wonderen, Europhys. Lett. \textbf{67}, 766 (2004).



\bibitem{suttorp-jpa}
L.G.~Suttorp, J. Phys. A: Math. Theor. {\bf 40}, 3697 (2007). 



\bibitem{boyd}
\newblock{R.W.Boyd, {\sl Nonlinear Optics}. 3rd ed. Academic Press, New York (2008).}

\bibitem{philbin-leonhardt} 
  T.~G.~Philbin, C.~Kuklewicz, S.~Robertson, S.~Hill, F.~Konig and U.~Leonhardt,
  Science {\bf 319}, 1367 (2008)
  [arXiv:0711.4796 [gr-qc]].



\bibitem{belgiorno-prl}
\newblock{F. Belgiorno, S.L. Cacciatori, M.Clerici, 
V. Gorini, G. Ortenzi, L. Rizzi, E. Rubino, V.G. Sala,  and D. Faccio, 
{\it Phys. Rev. Lett.} {\bf 105} 203901 (2010).}

\bibitem{rubino-njp}
\newblock{E.Rubino, F.Belgiorno, S.L.Cacciatori, M.Clerici, V.Gorini, G.Ortenzi, L.Rizzi, V.G.Sala, 
M.Kolesik, J.V.Moloney, D.Faccio, {\it New J. Phys.} {\bf 13} 085005 (2011).}

\bibitem{belgiorno-prd} 
  F.~Belgiorno, S.~L.~Cacciatori, G.~Ortenzi, L.~Rizzi, V.~Gorini and D.~Faccio,
  Phys.\ Rev.\ D {\bf 83}, 024015 (2011)
  [arXiv:1003.4150 [quant-ph]].



\bibitem{petev-prl} 
  M.~Petev, N.~Westerberg, D.~Moss, E.~Rubino, C.~Rimoldi, S.~L.~Cacciatori, F.~Belgiorno and D.~Faccio,
   Phys.\ Rev.\ Lett.\  {\bf 111}, 043902 (2013).


\bibitem{finazzi-carusotto-pra}
\newblock{S.Finazzi and I.Carusotto, {\it Phys. Rev.} {\bf A87}, 023803 (2013).}

\bibitem{finazzi-carusotto-pra14} 
  S.~Finazzi and I.~Carusotto,
  Phys.\ Rev.\ A {\bf 89}, 053807 (2014)
  [arXiv:1303.4990 [physics.optics]].

\bibitem{minkowski}
H.~Minkowski, Die Grundgleichungen f\"ur die elektromagnetischen Vorg\"{a}nge
in bewegten K\"{o}rpern. Nachrichten von der Gesellschaft der Wissenschaften zu G\"{o}ttingen, Mathematisch-Physikalische 
Klasse. S. 53–111 (1908).

\bibitem{post}
E.J.~Post, Formal structure of electromagnetics: general covariance and electromagnetics. North-Holland, 
Amsterdam (1962). 

\bibitem{Penfield-Haus}
\newblock{P.Penfield and H.A.Haus, {\sl Electrodynamics of moving media.} M.I.T. Press, Cambridge, Massachussetts
(1967).}

{\color{black}
\bibitem{belcacciadalla-hawking}
\newblock{F.Belgiorno, S.L.Cacciatori, F.Dalla Piazza,  
Phys.\ Rev.\ D {\bf 91}, 124063 (2015).}
}


\bibitem{gordon}
\newblock{W.Gordon, 
Ann. Phys. (Leipzig), 72, 421–456, (1923).}


\bibitem{balasz}
\newblock{N.L.Balazs, Jour. Optical Soc. Amer. {\bf 45}, 63 (1955).} 

\bibitem{phammauquan}
\newblock{Pham Mau Quan, Archives for Rational Mechanics and Analysis {\bf 1}, 54-80 (1957/58).}

\bibitem{synge}
\newblock{J.L. Synge, Geometrical optics in moving dispersive media. Commun.Dublin Inst.Ser.A 12 (1956).}

\bibitem{barcelo}
\newblock{C.Barcel\'o, S.Liberati, M.Visser, Living Rev. Relativity {\bf 14}, 3 (2011).}

\bibitem{horsley}
\newblock{S.A.R.Horsley, {\sl Phys. Rev. A} {\bf 86}, 023830 (2012).}



\bibitem{Gitman-Tyutin}
\newblock{D.M.Gitman and I.V.Tyutin, {\sl Quantization of Fields with Constraints.} Springer Series in Nuclear
and Particle Physics, Springer, Berlin (1990).}

\bibitem{Henneaux-Teitelboim}
\newblock{M.Henneaux and C.Teitelboim, {\sl Quantization of Gauge Systems.} Princeton University
Press, Princeton (1994).}


\bibitem{Watson-Jauch} 
  K.~M.~Watson and J.~M.~Jauch,
  Phys.\ Rev.\  {\bf 75}, 1249 (1949).


\bibitem{DeGroot-Suttorp}
\newblock{S.R.De Groot and L.G.Suttorp, {\sl Foundations of Electrodynamics.} Elsevier (1972).}

\bibitem{zuber}
\newblock{C.Itzykson and J-C.Zuber, {\sl Quantum Field Theory.} McGraw Hill (1980).}

{\color{black}
\bibitem{dirac-book}
\newblock{P.A.M. Dirac, {\sl Lectures on Quantum Mechanics}. Belfer Graduate School of Science, 
Yeshiva University. New York, 1964.}

\bibitem{dirac-papers}
\newblock{P.A.M. Dirac, Can. J. Math. {\bf 2}, 129 (1950); {\sl ibid}. {\bf 3}, 1 (1951).}
}

\bibitem{Rothe}
\newblock{Heinz J. Rothe and Klaus D. Rothe, {\sl Classical and Quantum Dynamics of Constrained 
Hamiltonian Systems.} World Scientific, Singapore (2010).}

\bibitem{barcelos-neto}
\newblock{J.Barcelos-Neto, Ashok Das and W.Scherer, {\sl Acta Phys. Pol.} {\bf B18} 269 (1987).}

{\color{black}
\bibitem{dimock-qm}
\newblock{J.Dimock, {\sl Quantum Mechanics and Quantum Field Theory}. Cambridge University Press, 
Cambridge (2011).}}

\bibitem{huttner-lett}
\newblock{B.Huttner,J.J. Baumberg and S.M.Barnett, {\sl Europhys. Lett.} {\bf 16}, 177 (1991).}

\bibitem{Nakanishi-Ojima}
\newblock{N.Nakanishi and I. Ojima, {\sl Covariant Operator Formalism of Gauge Theories and Quantum Gravity.}
Wolrd Scientific, Singapore (1990).}




{\color{black}
\bibitem{greiner-quantization}
\newblock{W.Greiner and J.Reinhardt, {\sl Field Quantization}.
Springer, Berlin (1996).}
}


\bibitem{eastham}
M.S.F. Eastham, {\sl The Asymptotic Solution of Linear Differential Systems. Applocations of the Levinson Theorem}. 
London Mathematical Society monographs. New series 4. Clarendon Press, Oxford (1989). 


\bibitem{manogue}
\newblock{C.A.Manogoue, {\sl Ann. Phys.} {\bf 181}, 261 (1988).}

\bibitem{cacciatori-perturb}
  F.~Belgiorno, S.~L.~Cacciatori and F.~Dalla Piazza,
  Eur.\ Phys.\ J.\ D {\bf 68}, 134 (2014)
  [arXiv:1402.2838 [quant-ph]].

\bibitem{dallapiazza-letter}
\newblock{F.Dalla Piazza et al., forthcoming.}

\bibitem{westerberg}
N.~Westerberg, S.~Cacciatori, F.~Belgiorno, F.~Dalla Piazza and D.~Faccio,
  New J.\ Phys.\  {\bf 16}, 075003 (2014)
  [arXiv:1403.5910 [gr-qc]].



\end{thebibliography}
\end{document}